\documentclass[prb,aps,twocolumn,showpacs,floats]{revtex4}
\usepackage{graphicx,amsmath,epsf}

\DeclareMathOperator{\cov}{cov}

\begin{document}
\title{A semiclassical theory of the Ehrenfest-time dependence of
  quantum transport in ballistic quantum dots}

\author{Piet W.\ Brouwer and Saar Rahav}

\affiliation{Laboratory of Atomic and Solid State Physics, Cornell
  University, Ithaca 14853, USA}

\begin{abstract}
We present a trajectory-based semiclassical calculation of the 
Ehrenfest-time dependence of the weak localization correction and the
universal conductance fluctuations of a ballistic quantum dot with
ideal point contacts. While the weak localization correction is
proportional to $\exp(-\tau_{\rm E}/\tau_{\rm D})$, where $\tau_{\rm
  E}$ and $\tau_{\rm D}$ are the dot's Ehrenfest time and dwell time,
respectively, the variance of the conductance is found to be
independent of $\tau_{\rm E}$. The latter is in agreement
with numerical simulations of the quantum kicked rotator [Tworzydlo
  {\em et al.}, Phys.\ Rev.\ B {\bf 69}, 165318 (2004) and Jacquod and
  Sukhorukov, Phys.\ Rev.\ Lett.\ {\bf 92}, 116801 (2004)].
\end{abstract}

\maketitle

\section{Introduction}

Weak localization and universal conductance fluctuations are
manifestations of quantum interference on quantum
transport.\cite{kn:beenakker1991b} How these effects 
appear in a ballistic quantum dot, coupled to
source and drain reservoirs via ideal point contacts, is believed to 
depend on the ratio of the Ehrenfest time and the
mean dwell time $\tau_{\rm D}$.\cite{kn:aleiner1996} 
The Ehrenfest time $\tau_{\rm E}$ is the time it takes for
two classical trajectories, initially a Fermi wavelength apart, to
diverge and be separated by a distance comparable to the system 
size.\cite{kn:larkin1968,kn:zaslavsky1981}
Trajectories need to remain inside the dot
during at least a time $\tau_{\rm E}$ in order to contribute to weak
localization or universal conductance fluctuations.

For the weak localization correction $\delta G$, semiclassical theory
predicts that $\delta G \propto \exp(-\tau_{\rm E}/\tau_{\rm
  D})$,\cite{kn:aleiner1996,kn:adagideli2003,kn:rahav2005,kn:brouwer2005c,kn:jacquod2006} 
which has been verified using accurate numerical simulations of the 
quantum kicked rotator\cite{kn:rahav2005,kn:jacquod2006} ---
a ``quantum dot'' with ``stroboscopic''
dynamics.\cite{kn:tworzydlo2004b} For universal conductance
fluctuations, numerical simulations 
found no Ehrenfest-time dependence of 
$\mbox{var}\, G$.\cite{kn:tworzydlo2004,kn:jacquod2004}
This remarkable result remained unexplained because of the absence of
a semiclassical theory of universal conductance fluctuations in
ballistic quantum dots. It is the goal of this article to report such a 
theory and offer a microscopic
explanation for the absence of Ehrenfest-time 
dependence of conductance fluctuations seen in the numerical
simulations.

Two theoretical approaches have been taken to
address quantum transport in ballistic quantum dots. One is the theory
of Aleiner and Larkin,\cite{kn:aleiner1996} which considers
quantum corrections to the ballistic analogue of the ``diffuson''
and ``cooperon'' propagators that play a central role in the 
diagrammatic perturbation theory for disordered conductors.
The other approach
is based on an expression relating the dot's
scattering matrix to a sum over classical trajectories
connecting the two point
contacts.\cite{kn:baranger1993} In all cases
where both approaches have been used to calculate the same
observable, the results have been the same.
In this article, we we use the trajectory-based approach.

In the trajectory-based
semiclassical approach, the dot's conductance $G$ is calculated from
the Landauer formula,
\begin{equation}
  G = \frac{2 e^2}{h} T,
  \label{eq:Landauer}
\end{equation}
where $T$ is the total transmission of the dot. The transmission $T$ is
expressed as a double sum over classical trajectories $\alpha$, $\beta$
that connect the two
point contacts,\cite{kn:jalabert1990}
\begin{equation}
  T = \frac{1}{(N_1 + N_2) \tau_{\rm D}} \sum_{\alpha,\beta}
  A_{\alpha} A_{\beta} e^{i (S_{\alpha} - S_{\beta})/\hbar}.
  \label{eq:Ssemi}
\end{equation}
Here $A_{\alpha}$ and $A_{\beta}$ are stability amplitudes and
$S_{\alpha}$ and $S_{\beta}$ are the classical actions of the two
trajectories $\alpha$ and $\beta$. The classical trajectories $\alpha$ 
and $\beta$ start with initial transverse momentum compatible with the
same mode $n$ in the left contact and end with transverse momentum 
compatible with the same mode $m$ in the right contact. The modes 
in each contact have quantized transverse momentum
\begin{equation}
  p_{\perp}(m) = \pm \pi \hbar m/W_{j},\ \ m=1,\ldots,N_{j},
\end{equation}
where $W_j$ is the width of the contact, $N_j$ the number of modes in
the contact, and the subscript $j=1,2$
refers to the left and right contacts, respectively.

\begin{figure}
\epsfxsize=0.95\hsize
\epsffile{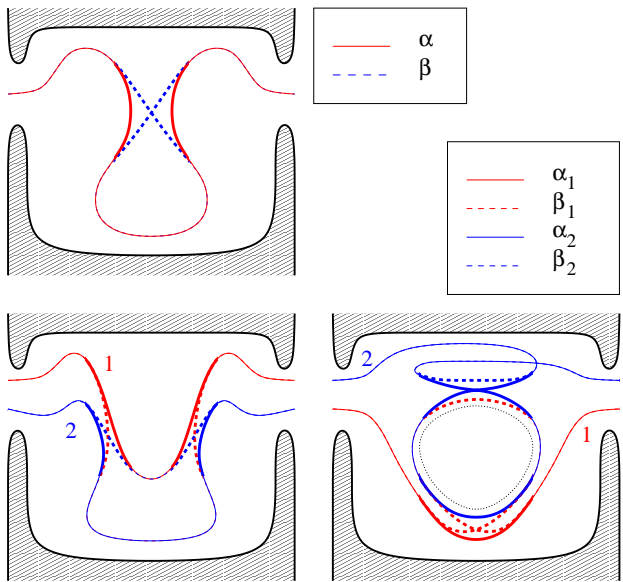}
\caption{\label{fig:00} 
  Top: schematic picture of a pair of interfering trajectories 
  $\alpha$ and $\beta$ that
  give rise to the weak localization correction to the conductance.
  Bottom left and bottom right: 
  schematic picture of quadruples of interfering trajectories
  $\alpha_1$, $\beta_1$, $\alpha_2$, and $\beta_2$
  that contribute to conductance fluctuations.
  The true 
  trajectories inside the dot are piecewise straight, with
  specular reflections off the dot's boundary. The small-angle (self)
  encounters are shown thick. In the bottom right 
  panel, the trajectories
  $\beta_1$ and $\alpha_2$ have one more revolution around a periodic
  trajectory (shown dotted) than their partners $\alpha_1$ and
  $\beta_2$.}
\end{figure}

Weak localization is the small negative correction $\delta G$ to the
ensemble average of the dot's conductance $G$ arising from quantum
interference. (The ensemble average is taken with respect to small
variations of the shape of the quantum dot or the Fermi energy.)
Two different trajectories $\alpha$ and $\beta$ contribute to
weak localization if their action difference $S_{\alpha} - S_{\beta}$
is of order $\hbar$. Pairs of trajectories $\alpha$ and $\beta$
that contribute to weak
localization are shown schematically in the top panel of
Fig.\ \ref{fig:00}. The two trajectories 
are almost equal, except for a stretch where $\beta$ has 
a small-angle self-intersection and $\alpha$ has a small-angle 
avoided self-intersection. Such pairs of
trajectories were originally pointed out by Aleiner and
Larkin;\cite{kn:aleiner1996} they were first investigated in the 
trajectory-based formalism by Sieber and Richter.\cite{kn:sieber2001}
The action difference between the two trajectories is of order $\hbar$
if the duration of the self-encounter is of order of the Ehrenfest
time. The probability that the trajectories do not escape through the
point contacts during the duration of the encounter is 
$\exp(-\tau_{\rm E}/\tau_{\rm D})$, hence the suppression $\delta G \propto 
\exp(-\tau_{\rm E}/\tau_{\rm D})$ of the weak localization correction
for large Ehrenfest times mentioned previously.\cite{kn:aleiner1996,kn:adagideli2003,kn:rahav2005,kn:jacquod2006,kn:brouwer2005c}

Since the transmission $T$ is a double sum over classical
trajectories $\alpha$ and $\beta$, the variance of the conductance is
expressed as a quadruple sum over classical trajectories $\alpha_1$,
$\beta_1$, $\alpha_2$, and $\beta_2$. Quadruples
of trajectories for which the individual action
differences $S_{\alpha_1} - S_{\beta_1}$ and $S_{\alpha_2} -
S_{\beta_2}$ are large (so that they do
not contribute to the average conductance), but the
total action difference $S_{\alpha_1} + S_{\alpha_2} - S_{\beta_1} -
S_{\beta_2}$ is small (of order $\hbar$) contribute to the conductance 
fluctuations. The two distinct 
classes of quadruples of trajectories that contribute to $\mbox{var}\,
G$ are shown in the bottom panels of Fig.\ \ref{fig:00}. Both classes
have their counterpart in the theory of conductance fluctuations in
diffusive conductors.\cite{kn:lee1985b,kn:altshuler1986}

The trajectories in the bottom
left panel of Fig.\ \ref{fig:00} are the ones commonly associated with
conductance fluctuations in ballistic quantum
dots\cite{kn:jalabert1990,kn:baranger1993b,kn:takane1998,kn:nakamura2004}
(see, however, Ref.\ \onlinecite{kn:argaman1996} for an exception).
They undergo two small-angle encounters. The
trajectories $\alpha_1$ and $\beta_1$, and $\alpha_2$ and $\beta_2$
are pairwise equal before the first encounter and after the last
encounter. Between the encounters $\alpha_1$ and
$\beta_2$, and $\alpha_2$ and $\beta_1$ are pairwise equal. The 
total action difference $S_{\alpha_1} + S_{\alpha_2} - S_{\beta_1} -
S_{\beta_2}$ is of order $\hbar$ if the duration of each encounter 
$\gtrsim \tau_{\rm E}$. Since the survival probability during each
encounter is $\exp(-\tau_{\rm E}/\tau_{\rm D})$, the contribution of
these trajectories to $\mbox{var}\, G$
decreases $\propto \exp(-2\tau_{\rm E}/\tau_{\rm D})$ for large
$\tau_{\rm E}$.

There is no exponential suppression at large Ehrenfest times
for the trajectories shown in the bottom right panel of Fig.\
\ref{fig:00}. Here, the trajectories $\beta_1$ and
$\alpha_2$ differ from $\alpha_1$ and $\beta_2$, respectively, by one
extra revolution around the same periodic trajectory (dotted). The
additional revolution around a periodic trajectory ensures that the 
individual action differences $S_{\alpha_1} - S_{\beta_2}$ and
$S_{\alpha_2} - S_{\beta_2}$ are large, whereas the total
action difference $S_{\alpha_1} + S_{\alpha_2} - S_{\beta_1} -
S_{\beta_2}$ is small if the trajectories $\alpha_1$ and
$\beta_2$ spend at least a time $\tau_{\rm E}$
near the periodic trajectory.
For a dot with mean dwell time
$\tau_{\rm D}$, the typical period of a periodic
trajectory (weighed with the square of the stability amplitudes) is
of order $\tau_{\rm D}$. This means that the trajectories $\alpha_1$,
$\beta_1$, $\alpha_2$, and $\beta_2$ need to wind many times around
the periodic trajectory if they are to contribute to
conductance fluctuations if $\tau_{\rm E} \gg \tau_{\rm D}$. However, 
since the survival
probability of the trajectories $\sim \exp(-\tau_{\rm p}/\tau_{\rm D})$
depends on the period $\tau_{\rm p}$ of the
periodic trajectory only, and not on the actual time $\sim
\tau_{\rm E}$ spent inside the quantum dot, the contribution to 
the conductance fluctuations is not
suppressed exponentially if $\tau_{\rm E} \gg \tau_{\rm D}$.

\begin{figure}
\epsfxsize=0.75\hsize
\epsffile{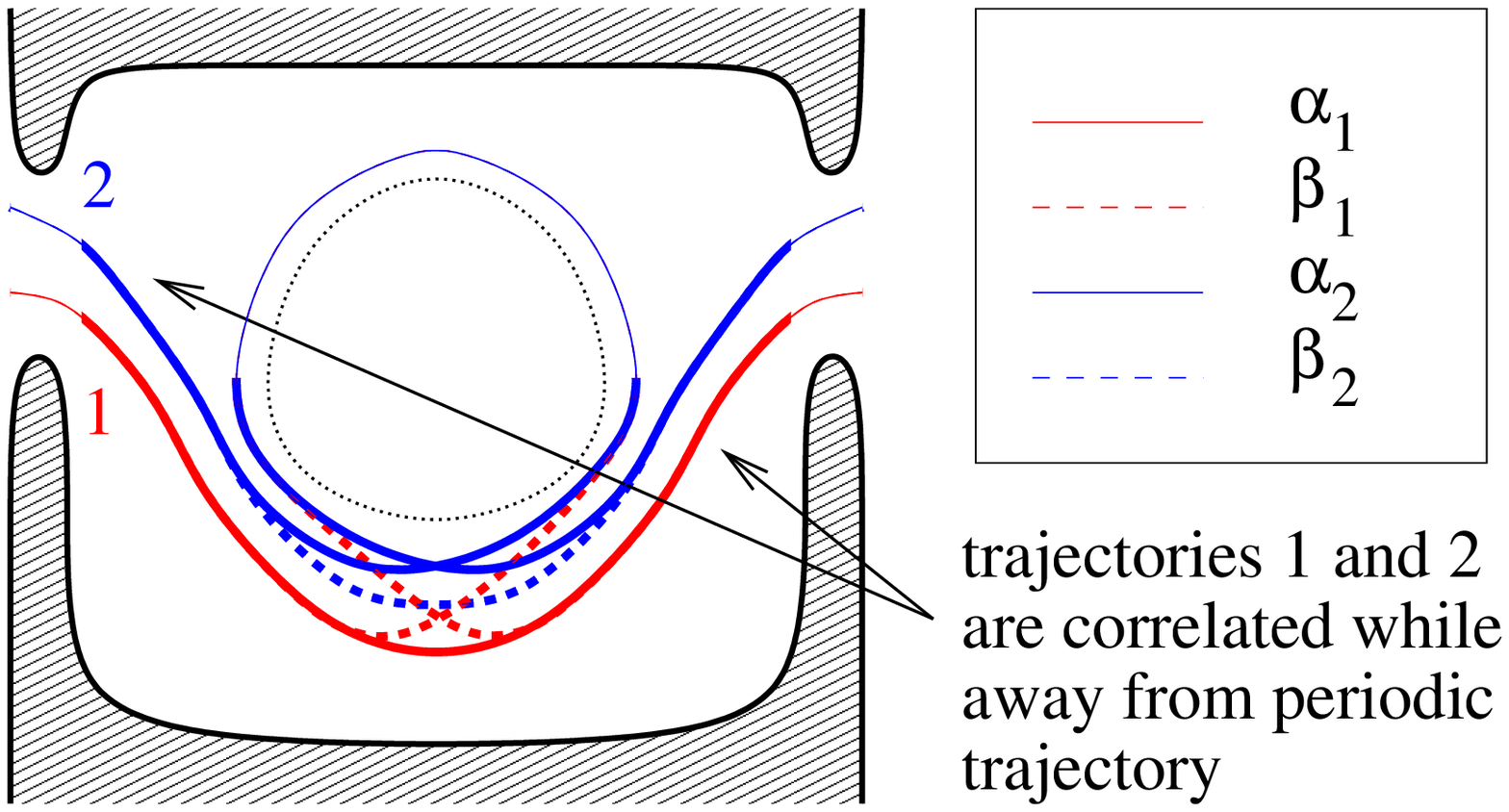}
\caption{\label{fig:0b} If the phase space distance between the
  trajectories $1$ and $2$ is of order $\hbar$ at the point when they
  arrive at or depart from the periodic trajectory, their motion is
  correlated for a time $\sim \tau_{\rm E}$ while away from the
  periodic trajectory. The contribution of such trajectories to 
  the conductance fluctuations is suppressed if $\tau_{\rm E} \gg
  \tau_{\rm D}$. After the removal of trajectories with correlated 
  motion away from the periodic trajectory the net contribution
  of the trajectories shown in the bottom right panel of
  Fig.\ \ref{fig:00} to the conductance
  fluctuations zero is finite.}
\end{figure}

This is not the full story, however. Periodic trajectories are
related to density of states fluctuations via Gutzwiller's trace
formula,\cite{kn:gutzwiller1990} so that trajectories of the type
shown in the bottom right panel of
Fig.\ \ref{fig:00} represent the impact of density-of-states
fluctuations on the conductance. However, for chaotic quantum dots
with ideal point contacts, the density of states and the conductance
are known to be statistically independent,\cite{kn:brouwer1997a} at least
in the regime $\tau_{\rm E} \ll
\tau_{\rm D}$ in which random matrix theory is valid.
In other words, there is no Einstein relation for the conductance of a
chaotic quantum dot with ideal point contacts if $\tau_{\rm E} \ll
\tau_{\rm D}$. The same conclusion can be drawn based on the
trajectory-based approach:
Indeed, if one sums the contribution of trajectories of the type shown
in the bottom right panel of
Fig.\ \ref{fig:00} over all initial and final positions where the
trajectories merge with or depart from the periodic 
trajectory, the net contribution to $\mbox{var}\, G$
vanishes.

The key
observation that explains the existence of Ehrenfest-time independent
conductance fluctuations in ballistic quantum dots is that the
cancellation that is responsible for the vanishing of the
density-of-states contribution to the conductance fluctuations 
is lifted if the dot's Ehrenfest time $\tau_{\rm E}$
is large in comparison to the mean dwell time $\tau_{\rm D}$. 
This is because correlations
between the trajectories shown in the bottom right panel of
Fig.\ \ref{fig:00} persist away from
the periodic trajectory if they
approach the periodic trajectory and/or leave it at very 
close phase space points, as shown in Fig.\ \ref{fig:0b}. 
These additional correlations, which last up to a time $\tau_{\rm
  E}$, suppress the contribution of 
these trajectories to the conductance fluctuations if $\tau_{\rm E} \gg
\tau_{\rm D}$, even if the two trajectories enter or exit through the
same contact. 
After removing the trajectories with these additional correlations
from the summation over all
trajectories of the type shown in the bottom right 
panel of Fig.\ \ref{fig:00}, the summation over 
the remaining trajectories no longer vanishes. 
The persistence of conductance fluctuations
if $\tau_{\rm E} \gg \tau_{\rm D}$ thus can be attributed to all
trajectories of the type shown in the bottom right panel of
Fig.\ \ref{fig:00} that do not arrive at
or depart from the periodic trajectory at close phase space
points.

These qualitative
arguments will be supported by the semiclassical calculations
presented in the next three sections. We use the trajectory-based
semiclassical formalism in the formulation developed in a series of 
works by Haake and 
collaborators.\cite{kn:mueller2004,kn:mueller2005,kn:heusler2006,kn:braun2005}
In Ref.\ \onlinecite{kn:heusler2006},
Heusler {\em et al.} show how this formalism is applied to the
calculation of the weak
localization correction, correcting two
canceling mistakes in the earlier theories of Refs.\
\onlinecite{kn:richter2002,kn:adagideli2003} (see also Refs.\
\onlinecite{kn:brouwer2005c,kn:jacquod2006}). Although developed for 
the limit $\tau_{\rm E} \ll \tau_{\rm D}$, the calculation of Ref.\
\onlinecite{kn:heusler2006} is readily extended to include
Ehrenfest-time dependences. The beginning of our
calculation follows that of Heusler {\em et al.}, but we take a
different classical limit at the end of the calculation.
Heusler {\em et al.} take the classical
limit $\hbar \to 0$ while keeping the number of
channels $N_1$ and $N_2$ in the two point contacts fixed. If the 
classical limit is taken this way, the ratio 
$\tau_{\rm E}/\tau_{\rm D} \to 0$, so that the Ehrenfest-time
dependence of the weak localization correction $\delta G$ and the
conductance variance $\mbox{var}\, G$ are lost. In order to 
preserve the Ehrenfest-time dependences of $\delta G$ and
$\mbox{var}\, G$, we take the limit
$\hbar \to 0$ while keeping the ratio $\tau_{\rm E}/\tau_{\rm D}$ 
fixed. For this classical limit, both the channel
numbers $N_1$ and $N_2$ and the dwell time $\tau_{\rm D}$
diverge, although the divergence of the dwell time is only logarithmic
in $\hbar$. The divergence of the channel numbers is of no concern 
for a calculation of
the weak localization correction or conductance fluctuations, 
since, if $N_1$ and $N_2$ are
large, both quantities depend on the ratio $N_1/N_2$ only. 
The divergence of the dwell time suppresses non-universal 
contributions to the quantum interference corrections. 
Finally, in this limit, interference of trajectories with 
more than one
small-angle self encounter (for weak localization) or with more 
than two small-angle encounters (for conductance fluctuations) 
can be neglected since the contribution of such trajectories 
to the average conductance is of order $1/(N_1+N_2)$ or
smaller.\cite{kn:heusler2006}

Before we
discuss our calculation of the Ehrenfest-time dependence of universal
conductance fluctuations, we review the trajectory-based calculation
of the weak localization correction to the conductance $G$. This
allows us to introduce the necessary formalism in a calculation that is
less complex than the calculation of conductance fluctuations.

\section{Weak localization}
\label{sec:2}


Without quantum
interference, the ensemble average $\langle G \rangle =
(2 e^2/h) N_1 N_2/(N_1+N_2)$, where $N_1$ and $N_2$ are the numbers
of propagating channels in the dot's left and right point contacts,
respectively. We are interested
in the small deviation $\delta G$
of $\langle G \rangle$ from its classical average.


\begin{figure}
\epsfxsize=0.95\hsize
\epsffile{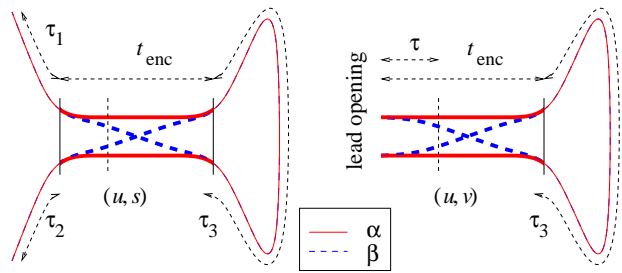}
\caption{\label{fig:1} 
Detail of the small-angle self-encounter and the definitions of
the various times used in the calculations. The beginning and end of
the encounter region are marked by solid thin lines, the location of
the Poincar\'e surface of section is marked by a dashed line. Each
trajectory passes the Poincar\'e surface of section twice. 
Left: Small-angle self encounter fully inside the dot. 
Right: Small-angle self-encounter that touches the lead opening. 
One end of the encounter region is marked by a solid thin line, 
the other end is the lead opening.
}
\end{figure}

Starting point of the calculation is the Landauer formula
(\ref{eq:Landauer}) together with the semiclassical expression
(\ref{eq:Ssemi}) for the dot's transmission. The pairs of classical
trajectories $\alpha$ and $\beta$
that contribute to weak localization are repeated
schematically in the left panel of Fig.\ \ref{fig:1}, together with
the definitions of the various times used in the calculation below.
Following Ref.\
\onlinecite{kn:heusler2006} (and referring there for details)
we find that the weak localization
correction to the transmission is equal to
\begin{eqnarray}
  \delta T &=& \frac{N_1 N_2}{(N_1 + N_2)^2} 
  \left( \prod_{j=1}^{2} \int_0^{\infty} \frac{d\tau_{j}}{\tau_{\rm
  D}} e^{-\tau_i/\tau_{\rm D}} \right)
  \nonumber \\ && \mbox{} \times
  \int_0^{\infty} d\tau_{3} e^{-\tau_3/\tau_{\rm D}}
  \int_{-c}^c ds du \frac{e^{i s u/\hbar -t_{{\rm enc}}/\tau_{\rm
  D}}}{2 \pi \hbar t_{{\rm enc}}}.~~
  \label{eq:F1}
\end{eqnarray}
Here $\tau_1$ and $\tau_2$ are the durations of the parts of the
trajectories between the encounter and the lead openings and $\tau_3$
is the duration of the loop, see Fig.\ \ref{fig:1}. The prefactor
$N_1 N_2/(N_1+N_2)^2$ consists of a factor $N_1$ from the free sum over 
the incoming channel in the left contact, a factor $N_2/(N_1+N_2)$
from the probability to escape through the right contact, 
and a factor $1/(N_1+N_2)$ 
from the small-angle phase space encounter.\cite{kn:heusler2006} 
The phase space
coordinates $u$ and $s$ parameterize the distance between the two 
parts of one of the 
interfering trajectories at a reference Poincar\'e surface of section
during the self-encounter, along unstable and stable directions in
phase space, respectively, see Fig.\ \ref{fig:1}.
The product $su$ in the exponent of Eq.\ (\ref{eq:F1}) is the
difference of the actions of the trajectories $\alpha$ and 
$\beta$.\cite{kn:spehner2003,kn:turek2003} 
The integration domain
is the set of phase space coordinates $s$ and $u$ for which the
distance between the two trajectories is smaller than a
cut-off $c$, {\em i.e.,} $|s|$, $|u| < c $. The cut-off $c$ is
a classical scale chosen small enough that the 
classical dynamics can
be linearized on phase space distances below $c$. 
The precise value of the cut-off
is not relevant in the limit $\hbar \to 0$.
The time $t_{{\rm enc}}$ is the duration of
the encounter, {\em i.e.}, the time
that the two segments of the trajectories $\alpha$ and $\beta$ 
are within a phase-space distance $c$,
\begin{equation}
  t_{{\rm enc}}
  = \frac{1}{\lambda} \ln \frac{c}{|s|} + \frac{1}{\lambda} \ln
  \frac{c}{|u|},
\end{equation} 
where $\lambda$ is the Lyapunov exponent of the classical motion inside
the quantum dot. It appears in the denominator in Eq.\ (\ref{eq:F1}) 
in order to cancel a
spurious contribution to the integral from the freedom to choose
the reference point inside the encounter 
region.\cite{kn:mueller2004,kn:mueller2005} 
The factors 
$\exp(-t_{{\rm enc}}/\tau_{\rm D})$ and $\exp(-\tau_3/\tau_{\rm D})$
are the probabilities not to escape during the encounter and during
the loop segment, respectively. Although the trajectory traverses the 
encounter twice, the escape probability involves only one encounter 
duration.\cite{kn:rahav2005,kn:heusler2006} 

The calculation of $\delta T$ closely follows the general
principle outlined in appendix D3 of Ref.\
\onlinecite{kn:mueller2005}. Limiting the integration domain to
positive $u$, we first rewrite Eq.\ (\ref{eq:F1}) as
\begin{equation}
  \delta T = \frac{2 N_1 N_2 \tau_{\rm D}}{(N_1 + N_2)^2} 
  \int_0^{c} du \int_{-c}^{c} ds
    \frac{e^{-t_{{\rm enc}}/\tau_{\rm D}} 
  \cos(su/\hbar)}{2 \pi \hbar t_{{\rm enc}}}.
\end{equation}
We then perform the variable change 
\begin{equation} 
  u = c/\sigma,\ \  s = c x \sigma.
  \label{eq:varchg1}
\end{equation}
With the new integration variables, the integration domain is $-1 < x <
1$ and $1 < \sigma < 1/|x|$. Further, 
$t_{{\rm enc}} = \lambda^{-1} \ln(1/|x|)$. Hence, with $r = c^2/\hbar$,
the integral (\ref{eq:F1}) becomes
\begin{eqnarray}
  \delta T &=& \frac{N_1 N_2 r \tau_{\rm D}}{(N_1 + N_2)^2} 
  \int_{-1}^{1} dx
  \frac{\cos (x r) e^{ - t_{{\rm enc}}/\tau_{\rm D}}}{\pi
  t_{{\rm enc}}} 
  \int_{1}^{1/|x|}  \frac{d\sigma}{\sigma}
  \nonumber \\
  &=&
  \frac{2 r \lambda \tau_{\rm D} N_1 N_2}{(N_1 + N_2)^2\pi} 
  \int_0^{1} dx x^{1/\lambda \tau_{\rm D}} \cos (x r)
  \nonumber \\ &=&
  \frac{2 N_1 N_2}{(N_1 + N_2)^2 \pi} 
  \left[ \lambda \tau_{\rm D} \sin r
  \vphantom{\int}
  \right. \nonumber \\ && \left. \mbox{}
  - r^{-1/\lambda \tau_{\rm D}}
  \int_0^{r} dx x^{1/\lambda \tau_{\rm D}} 
  \frac{\sin x}{x} \right].
  \label{eq:F1calc}
\end{eqnarray}
The term proportional to $\sin r$ is a rapidly
oscillating function
of Planck's constant and is discarded in the classical limit.
Writing the remaining term in terms of the Ehrenfest time,
\begin{equation}
  \tau_{\rm E} = \frac{1}{\lambda} \ln r =
  \frac{1}{\lambda} \ln \frac{c^2}{\hbar}.
  \label{eq:tauE}
\end{equation}
we find
\begin{equation}
  \delta T = - \frac{N_1 N_2}{(N_1 + N_2)^2} \frac{2}{\pi} 
  e^{-\tau_{\rm E}/\tau_{\rm D}}
  \int_0^{r} dx x^{1/\lambda \tau_{\rm D}} \frac{\sin x}{x}.
  \label{eq:F1intermediate}
\end{equation}
The limit $\hbar \to 0$ at a fixed ratio of
Ehrenfest time and dwell time consists of sending both 
$r = c^2/\hbar \to \infty$ and $\lambda \tau_{\rm D} \to \infty$. These
limits can be taken independently in Eq.\ (\ref{eq:F1intermediate}).
We then find
\begin{equation}
  \delta T = -\frac{N_1 N_2}{(N_1 + N_2)^2} 
  e^{-\tau_{\rm E}/\tau_{\rm D}}. \label{eq:F1res}
\end{equation}
The exponential 
dependence of Eq.\ (\ref{eq:F1res}) is in agreement with previous
calculations of the Ehrenfest-time dependence of weak
localization.\cite{kn:adagideli2003,kn:rahav2005,kn:brouwer2005c,kn:jacquod2006} Note that the
appearance of the classical phase-space cut-off $c$ in the definition
of the Ehrenfest time does not affect the final result: In the limit
$\hbar \to 0$ at fixed $\tau_{\rm E}/\tau_{\rm D}$, the dwell time
$\tau_{\rm D} \to \infty$, which removes any $c$-dependence from the
final expressions.

Instead of calculating the weak localization correction to the
transmission, one may also calculate the quantum interference
correction to the ensemble averaged reflection. Unitarity relates 
the total reflection $R_i$ off contact $i$ to the total transmission 
$T$, 
\begin{equation}
  R_i = N_i - T,\ \ i=1,2.
\end{equation}
Although this implies $\delta R_1 = \delta R_2 = - \delta T$, 
it remains instructive to verify this result explicitly from the
semiclassical formalism. 

The semiclassical formula for the total reflection $R_i$ is the same
as Eq.\ (\ref{eq:Ssemi}) for the total transmission $T$, but with a
double sum over trajectories that connect contact $i$ to itself.
The quantum correction to $R_i$ consists of two parts: the counterpart
$\delta R_i^{(1)}$
of the weak localization correction to the transmission $T$, which
involves encounters in the interior of the quantum dot, and an extra
quantum correction $\delta R_i^{(2)}$
from encounters that touch the lead opening. The calculation of 
the first correction to $R_i$ 
only differs from the calculation of $\delta T$ in the replacement of
$N_1 N_2$ by $N_i^2$, $i=1,2$, so that
\begin{eqnarray}
  \delta R_{i}^{(1)} &=&
  -\frac{N_i^2}{(N_1 + N_2)^2} 
  e^{-\tau_{\rm E}/\tau_{\rm D}}.
  \label{eq:Ri1}
\end{eqnarray}

The calculation of the reflection correction 
$\delta R_i^{(2)}$ from encounters that
touch the lead opening is a little
different.\cite{kn:rahav2006,kn:rahav2006b,kn:jacquod2006} This
correction is usually referred to as the `coherent backscattering'
correction to reflection. In the limit $\tau_{\rm E} \ll \tau_{\rm
  D}$, coherent backscattering can be calculated using the `diagonal
approximation' for the double sum over trajectories in Eq.\
(\ref{eq:Ssemi}).\cite{kn:doron1991,kn:lewenkopf1991,kn:baranger1993} 
In the diagonal approximation, only 
trajectories $\alpha$ and $\beta$ that
are identical up to time reversal are
kept. If
$\tau_{\rm E} \gtrsim \tau_{\rm D}$, the diagonal approximation fails,
however, and a full summation over families of
trajectories very similar to the trajectory sums
for weak localization is called for.\cite{kn:rahav2006b,kn:jacquod2006}
%
%

Since the trajectories $\alpha$ and $\beta$ have the same
perpendicular component of the momentum at the lead opening upon
entrance as well as upon exit, the
optimal choice of phase space coordinates for an encounter that 
touches the lead opening is a 
phase space coordinate $v$ that represents the perpendicular momentum at the
contact, together with the unstable phase space direction $u$
(taken with respect to motion away from the contact). 
Note that $v$ is a meaningful coordinate for a Poincar\'e surface of
section in an encounter that touches the lead opening, because all
trajectories piercing the Poincar\'e surface of section exit the
quantum dot together.
The stable phase space coordinate $s$, which was used for
small-angle self-encounters that contribute to weak
localization, is not a good choice here, because in general
the trajectories $\alpha$
and $\beta$ have different $s$ if the self encounter touches the lead
opening. Following Refs.\
\onlinecite{kn:spehner2003,kn:turek2003}, we normalize $v$ such that 
the cross section volume element in
phase space is $du dv$. With this normalization, the phase space coordinates
$(u,v)$ are uniformly distributed for ergodic motion. As in the case
of the calculation of the weak localization correction $\delta T$, we 
use the first passage of the trajectory $\alpha$
through the encounter region 
as a reference, consider a Poincar\'e surface of section at an arbitrary
point during the encounter, and denote the phase space coordinates at 
which $\alpha$ cuts through the Poincar\'e surface of section a
second time by $(u,v)$. The action 
difference $\Delta S$ is the phase space area enclosed by the four
segments of the trajectories $\alpha$ and $\beta$
involved in the interference
correction.\cite{kn:spehner2003,kn:turek2003}
Recalling that $\alpha$ and its partner trajectory  $\beta$
have perpendicular momenta compatible with the same modes in the lead
opening, the phase space coordinates of the first and second passages 
of $\beta$ are $(u,0)$ and $(0,v)$, respectively, hence
$\Delta S = uv$. Note that the enclosed phase space areas are
conserved along the motion of the trajectories. In particular, this
means that the action difference $\Delta S$ is independent of where the
Poincar\'e surface of section is chosen and that 
the coordinate $v$ scales $\propto \exp(-\lambda t)$ upon
moving away from the contacts.

With this, we find that the coherent
backscattering correction reflection becomes
\begin{eqnarray}
  \delta R_{i}^{(2)} &=&\frac{N_i}{(N_1 + N_2)} 
  \int d\tau_3 e^{-\tau_3/\tau_{\rm D}}
  \int_{-c}^{c} dv du
  \nonumber \\ && \mbox{} \times
  \int_0^{\lambda^{-1} \ln(c/|v|)}
  \frac{d\tau}{\tau_{\rm D}} 
  \frac{e^{i v u/\hbar -t_{\rm enc}/\tau_{\rm
  D}}}{2 \pi \hbar t_{\rm enc}},
  \label{eq:Ri2sub}
\end{eqnarray}
where $\tau$ is the time needed to go between the Poincar\'e surface of 
section and the lead opening, see Fig. \ref{fig:1}. The prefactor
$N_i$ arises from the summation over the incoming transverse
channel. Note that there is no additional factor $N_i/(N_1+N_2)$ [as
in the case of an encounter that resides in the interior of the
quantum dot, {\em cf.}\ Eq.\ (\ref{eq:F1})], because the probability of 
escape
through contact $i$ is unity for encounters that touch the lead
opening, $i=1,2$. In Eq.\ (\ref{eq:Ri2sub}) the 
encounter time is a function of $\tau$ and $u$,
\begin{equation}
  t_{\rm enc}(\tau,u) = \tau + \lambda^{-1} \ln(c/|u|).
\end{equation}
The condition $\tau < \lambda^{-1} \ln(c/|v|)$ in Eq.\
(\ref{eq:Ri2sub}) ensures that the encounter touches the lead 
opening. (The precise value of the cut-off $c$ is unimportant in the
limit $\hbar \to 0$, as before.)
In order to perform the integrations in Eq.\
(\ref{eq:Ri2sub}), we take $u$ to be positive and perform the 
variable change 
\begin{equation} 
  \tau' = \tau + \lambda^{-1} \ln(c/|u|), \ \
  u = c/\sigma,\ \  v = c x \sigma.
  \label{eq:varchg2}
\end{equation}
With the new integration variables, the integration domain is $-1 < x <
1$, $1 < \sigma < e^{\lambda \tau'}$, and $0 < \tau' < \lambda^{-1}
\ln(1/|x|)$. Further, $t_{{\rm enc}} = \tau'$. Hence, with 
$r = c^2/\hbar$, the integral (\ref{eq:Ri2sub}) becomes
\begin{eqnarray}
  \delta R_i^{(2)} &=& \frac{N_i \tau_{\rm D} r}{\pi(N_1
  + N_2)} \int_{-1}^{1} dx \int_0^{\lambda^{-1} \ln(1/|x|)} 
  \frac{d\tau'}{\tau_{\rm D}} e^{-\tau'/\tau_{\rm D}}
  \nonumber \\ && \mbox{} \times
  \int_1^{e^{\lambda \tau'}} \frac{d\sigma}{\sigma \tau'}
  \cos(r x) \nonumber \\ &=&
  \frac{2 N_i \lambda \tau_{\rm D} r}{\pi(N_1
  + N_2)} \int_{0}^{1} dx 
  (1 - x^{1/\lambda \tau_{\rm D}}) \cos(rx).
  \nonumber \\
\end{eqnarray}
Neglecting terms proportional to $\sin r$, the remaining integrals are
the same as for the calculation of $\delta T$, and one finds the
result\cite{kn:rahav2006,kn:rahav2006b,kn:jacquod2006}
\begin{equation}
  \delta R_{i}^{(2)} =
  \frac{N_i}{N_1+N_2} e^{-\tau_{\rm E}/\tau_{\rm D}}.
  \label{eq:Ri2}
\end{equation}
Adding Eqs.\ (\ref{eq:Ri1}) and (\ref{eq:Ri2}), one finds $\delta R =
-\delta T$, as is required by unitarity.

\section{Conductance Fluctuations}
\label{sec:3}

The same theoretical framework can be used to calculate universal
conductance fluctuations. Technically, it is most convenient to
calculate the covariance of reflection coefficients $R_1$ and $R_2$
for reflection from the left and right point contacts, because it
avoids the necessity of dealing with encounters that touch the two
point contacts. (The case of encounters that touch the lead openings
will be discussed at the end of this section.)
The covariance of the reflection coefficients is
directly related to the conductance variance,
\begin{equation}
  \mbox{var}\, G =
  \left( \frac{2 e^2}{h} \right)^2 \mbox{var}\, T =
  \left( \frac{2 e^2}{h} \right)^2 \mbox{cov}\, (R_1, R_2).
\end{equation}
\begin{figure}
\epsfxsize=0.95\hsize
\epsffile{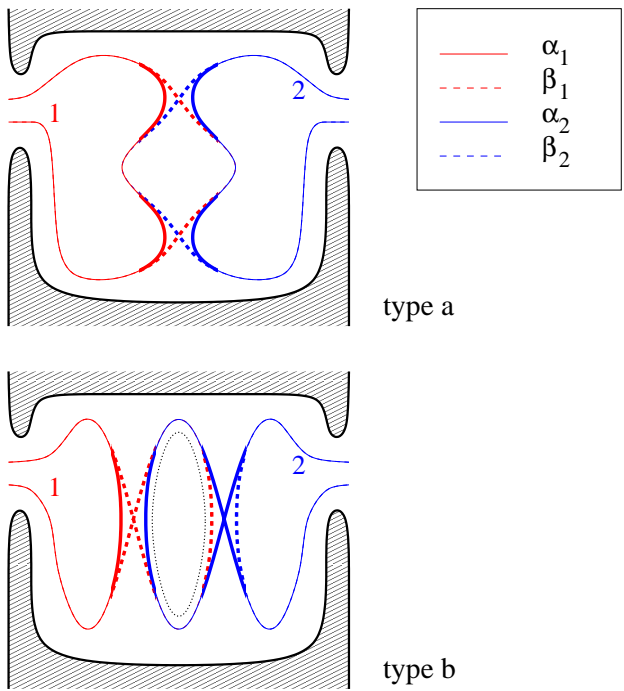}
\caption{\label{fig:2} 
  Two configurations of interfering trajectories that contribute to 
  the covariance $\mbox{cov}\, (R_1,R_2)$ of 
  reflection coefficients for the left and right contacts. The
  encounter regions, segments of the trajectories
  for which the phase space distance between the
  trajectories $1$ and $2$ is less than a classical cut-off $c$, are
  shown thick. In the top configuration, the trajectories $1$ and $2$
  have two consecutive well-separated encounters. In the bottom panel,
  the trajectories $\alpha_2$ and $\beta_1$ have one more revolution
  around a periodic trajectory (shown dotted) than their counterparts
  $\alpha_1$ and $\beta_2$.}
\end{figure}

\begin{figure}
\epsfxsize=0.95\hsize
\epsffile{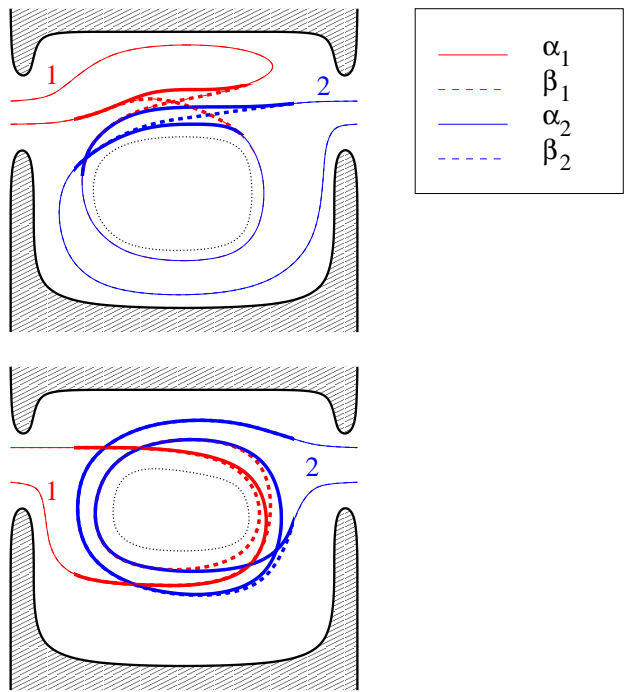}
\caption{\label{fig:3} 
Two examples of interfering trajectories with overlapping
encounters. The encounter regions are shown thick. In both panels, 
the trajectories $\beta_1$ and $\alpha_2$ have one more 
revolution around a closed periodic trajectory
(shown dotted) than their partners $\alpha_1$ and $\beta_2$.
The top panel shows two encounters that overlap at one end. This
is the ``three encounter'' of Ref.\
\onlinecite{kn:heusler2006}. The bottom panel shows two encounters
overlapping at both ends, so that the encounter region forms a closed
loop. Although encounters that themselves form a closed loop do not 
contribute to the conductance variance if $\tau_{\rm E} \ll \tau_{\rm
  D}$, their contribution is essential if $\tau_{\rm E} \gg
\tau_{\rm D}$.}
\end{figure}

The reflections $R_1$ and $R_2$ are expressed in terms of a double
sum over
classical trajectories $\alpha_i$ and $\beta_i$, $i=1,2$,
connecting each contact to itself, similar to
the trajectory sum of Eq.\ (\ref{eq:Ssemi}) for the dot's
transmission. The covariance $\mbox{cov}\, (R_1,R_2)$ then becomes a
quadruple sum over classical trajectories $\alpha_1$, $\beta_1$,
$\alpha_2$, and $\beta_2$. There are two distinct configurations of four 
interfering trajectories 
that contribute to $\mbox{cov}\, (R_1, R_2)$, see Fig.\ \ref{fig:2}. 
These are the
counterparts for reflection of the trajectories shown in the two
bottom panels of Fig.\ \ref{fig:00}. We refer to them as ``type a'' and
``type b'' interfering trajectories.\footnote{A previous version of
this article, Ref.\ \onlinecite{kn:brouwer2005c}, distinguished three
configurations of interfering trajectories. The first of these is
what is called type a here. The second and third categories of
Ref.\ \onlinecite{kn:brouwer2005c} are both of type b using the
classification of the present article. However, Ref.\ 
\onlinecite{kn:brouwer2005c} omits essential other configurations 
of interfering trajectories that are also contained in type b, 
such as the configuration shown in the bottom panel of Fig.\ \ref{fig:3}.}
Other possible configurations of interfering trajectories
will give contributions to $\mbox{var}\, G$ smaller by a power of
$N_1+N_2$ and need not be considered in the limit $\hbar \to 0$ at
fixed $\tau_{\rm E}/\tau_{\rm D}$ we
consider here.\cite{kn:heusler2006} 
The configuration of Fig.\ \ref{fig:2}b is not
left-right symmetric and acquires an extra factor two. 
Denoting the contributions from interfering trajectories of types a
and b as $A$ and $B$, respectively, we then have
\begin{equation}
  \mbox{var}\, G = 2 \left( \frac{2 e^2}{h} \right)^2
  (A + 2 B).
  \label{eq:varGF}
\end{equation}
The prefactor $2$ in Eq.\ (\ref{eq:varGF}) only appears in the
presence of time-reversal symmetry. It accounts for the configurations
of interfering trajectories obtained by time-reversing the trajectories
originating from the left contact but not the trajectories originating
from the right contact.

The primary distinction between the types a and b is that the 
trajectories of the former type do not have segments that are 
close to a periodic trajectory, whereas in the latter case
they do. (The periodic trajectory is the dotted loop in the right
panel of Fig.\ \ref{fig:2}; Here ``close'' means that a segment of the
trajectories $\alpha$ or $\beta$ can be deformed into a periodic
trajectory within the region of phase space in which the chaotic
dynamics in the quantum dot can be linearized.) This means that $B$ contains
all conductance fluctuations that are tied to density of states
fluctuations, whereas $A$ represents the density-of-states
independent fluctuations of the conductance. In the limit $\tau_{\rm
  E} \ll \tau_{\rm D}$ the conductance and the density of states are
statistically independent,\cite{kn:brouwer1997a} so that we expect
that $B=0$ in that limit.

Although the encounters as drawn 
in Fig.\ \ref{fig:2} do not overlap, they can overlap in
principle. In fact, such overlaps are essential for a theory of universal
conductance fluctuations if the Ehrenfest time is larger than the
dwell time and, hence, comparable to the total path length.
Two examples of
overlapping encounters are shown in Fig.\ \ref{fig:3}. The top panel
of Fig.\ \ref{fig:3} shows the ``three-encounter'' of Ref.\
\onlinecite{kn:heusler2006}. The bottom panel shows a more complicated
configuration in which the two encounters overlap both at their
beginning and at their end, so that the encounter region forms a
closed loop. In principle, overlapping encounters 
can arise from bringing the two 
encounters of Fig.\ \ref{fig:2}a close together along one of the 
two solid trajectories. However, as soon as the encounters of Fig.\ 
\ref{fig:2}a overlap, one of the trajectories involved has a segment
close to a periodic trajectory. Therefore, according to our definitions of
the covariance contributions $A$ and $B$, the two 
encounters in the configuration of type a do not overlap.

Calculation of the contribution $A$ of interfering trajectories of 
type a is straightforward since the quantum interference correction 
from non-overlapping encounters factorizes,\cite{kn:heusler2006} 
\begin{equation}
  A = \frac{(N_1 N_2)^2}{(N_1 + N_2)^4} e^{-2 \tau_{\rm
  E}/\tau_{\rm D}}.
\end{equation}

\begin{figure}
\epsfxsize=0.85\hsize
\epsffile{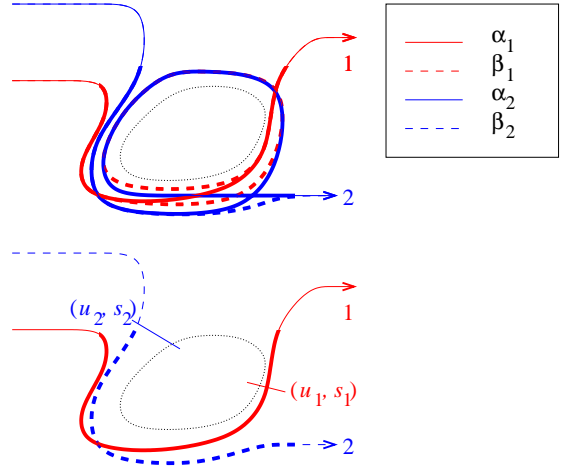}
\caption{\label{fig:4} 
  Top: Interfering trajectories of type b (dashed and solid
  curves) and the periodic trajectory (dotted).
  Bottom: Short versions of the interfering trajectories together with
  the Poincar\'e surfaces of section. At each Poincar\'e surface of
  section the stable and unstable phase space coordinates $s$ and $u$
  parameterize the distance to the periodic trajectory.}
\end{figure}

The calculation of the contribution of interfering trajectories of
type b significantly more involved. Because of the existence of
configurations as shown in the bottom panel of Fig.\ \ref{fig:3}, in
which the encounter region winds one or several times around a periodic
trajectory, one cannot calculate the contribution from trajectories of
type b by considering two-encounters and three-encounters
only, as was done in a previous version of this
article.\cite{kn:brouwer2005c} (Note, however, that since 
encounters that fully wind around a periodic trajectory with period
$\tau_{\rm p}$ 
exist for periods $\tau_{\rm p} \lesssim \tau_{\rm E}$ only, considering 
two-encounters and three-encounters only is sufficient if
$\tau_{\rm E} \ll \tau_{\rm D}$.\cite{kn:heusler2006} See Ref.\
\onlinecite{kn:mueller2005} for a precise
calculation that verifies this for closed quantum dots.)

In order to parameterize the
combinations of interfering trajectories of type b, we use coordinates
that measure the phase space distance to the periodic trajectory. 
The periodic trajectory, as well
as the four interfering trajectories of type b are shown again in the top 
panel of Fig.\
\ref{fig:4}. The interfering trajectories consist of two
``short'' trajectories (trajectories $\alpha_1$ and $\beta_2$ in Fig.\
\ref{fig:2} or Fig.\ \ref{fig:4}) 
and two ``long'' trajectories (trajectories $\alpha_2$ and
$\beta_1$ in Fig.\ \ref{fig:2} or Fig.\ \ref{fig:4}), where the long
trajectories have one more revolution around the periodic trajectory than
the short one. We use the two short trajectories as our reference. At a
point where the phase space distance to the periodic
trajectory is less than the classical cut-off $c$, for each short 
trajectory $i=1,2$
we draw a Poincar\'e surface of section and use phase space
coordinates $(u_i,s_i)$, $i=1,2$, referring to the unstable and stable
directions in phase space to describe the distance to the periodic
trajectory, see Fig.\ \ref{fig:4}b. Notice that we need to
draw two separate Poincar\'e surfaces of section because the two
short trajectories do not need to be within a phase space distance $c$
from the periodic trajectory at the same time. The long
versions of the trajectories pass through these Poincar\'e surfaces of
section twice and have phase space coordinates $(u_i,s_i
e^{-\lambda \tau_{\rm p}})$ or $(u_i e^{-\lambda \tau_{\rm p}},s_i)$,
$i=1,2$, where $\tau_{\rm p}$ is the period of the periodic trajectory. 

\begin{figure}[t]
\epsfxsize=0.95\hsize
\epsffile{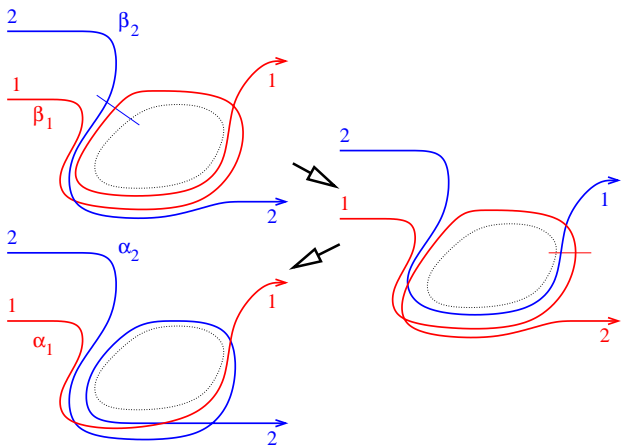}
\caption{
  \label{fig:5}
  Two successive deformations used to calculate the action difference
  between the pair of trajectories in the top left diagram and the
  pair of trajectories in the bottom left diagram.}
\end{figure}

In order to calculate the action difference $\Delta S = S_{\alpha_1} +
S_{\alpha_2} - S_{\beta_1} - S_{\beta_2}$
we perform two successive deformations, as
shown in Fig.\ \ref{fig:5}. The action difference for the first
deformation can be calculated using the Poincar\'e surface of section
drawn in the top left panel of Fig.\ \ref{fig:5}. The phase space
coordinates of the two trajectories are $(u_2,s_2)$ and 
$(u_1 e^{-\lambda (\tau_{\rm p}-\tau')},s_1 e^{-\lambda \tau'})$, where
$\tau'$ is the time difference between the two Poincar\'e 
sections in the bottom panel of Fig.\ \ref{fig:4}, measured along the
periodic trajectory. Then the corresponding action
difference is $(u_2 - u_1 e^{-\lambda (\tau_{\rm p}-\tau')})(s_2 - s_1
e^{-\lambda \tau'})$, see Refs.\ 
\onlinecite{kn:spehner2003,kn:turek2003}. The
action difference for the second deformation is calculated using the
Poincar\'e surface of section drawn in the right panel of Fig.\
\ref{fig:5}. The phase space coordinates of the two trajectories involved
here are $(u_1,s_2 e^{-\lambda(\tau_{\rm p}-\tau')})$ and $(u_2
e^{-\lambda \tau'},s_1)$, corresponding to the action difference $(u_1
- u_2 e^{-\lambda \tau'})(s_2 e^{-\lambda(\tau_{\rm p}-\tau')} - s_1)$. Adding
  the two action differences, one finds
\begin{equation}
  \Delta S = (u_2 s_2 - u_1 s_1)(1 - e^{-\lambda \tau_{\rm p}}).
  \label{eq:dS}
\end{equation}
The sign of the action difference is not relevant, as both $\Delta S$
and $-\Delta S$ appear in the final summation. Our expression for the
action difference $\Delta S$ differs by
a factor $1 - e^{-\lambda \tau_{\rm p}}$
from the action difference used in Ref.\
\onlinecite{kn:heusler2006}. This difference is unimportant, since
relevant periods $\tau_{\rm p}$ are of the order of the mean dwell time
$\tau_{\rm D}$ and $\exp(-\lambda \tau_{\rm D}) \ll 1$.

We count periodic trajectories that consist of several revolutions
of one shorter trajectory as separate trajectories. This correctly
takes into account the contribution from interfering trajectories where
the difference between interfering trajectories is more than one
revolution around a periodic trajectory.

\begin{figure}[t]
\epsfxsize=0.6\hsize
\epsffile{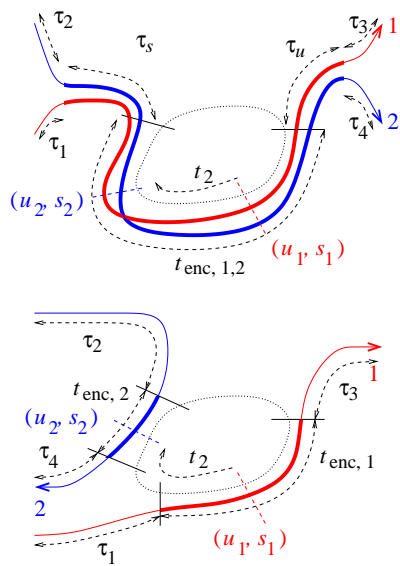}
\caption{
  \label{fig:6}
  Definitions of the times appearing in Eq.\ (\ref{eq:B}). The top
  panel shows an example where the two short trajectories $\alpha_1$
  and $\beta_2$ (solid curves) have strong
  correlations before and after arriving within the vicinity of the
  periodic trajectory (dotted curve). The time during which
  these strong classical correlations exist is $\tau_s$ or
  $\tau_u$. The bottom panel shows an example where there are no
  classical correlations except for the time the trajectories are in
  the vicinity of the periodic trajectory. In this case,
  $\tau_s = \tau_u = 0$. In the bottom panel, the two trajectories
  have different encounter times $t_{{\rm enc},1}$ and $t_{{\rm enc},
  2}$. In the example shown in the top panel, the two encounter times
  $t_{{\rm enc},1}$ and $t_{{\rm enc},
  2}$ are (almost) equal. The thin solid lines indicate the beginning and
  the end of the encounter with the periodic trajectory; the total
  encounter regions are marked by thick trajectories. The
  dashed lines indicate the positions where the Poincar\'e surfaces of
  section for the two trajectories are taken.}
\end{figure}

Now we are ready to calculate the contribution $B$ of trajectories of
type b to the reflection covariance. Repeating the steps used for the
calculation of the weak localization correction $\delta T$, we find
\begin{eqnarray}
  B &=& \frac{N_1^2 N_2^2}{(N_1+N_2)^4} 
  \left( \prod_{j=1}^{4} \int_0^{\infty} \frac{d\tau_{j}}{\tau_{\rm D}}
  e^{-\tau_j/\tau_{\rm D}} \right)
  \int d\tau_{\rm p} e^{-\tau_{\rm p}/\tau_{\rm D}} 
  \nonumber \\ && \mbox{} \times
  \int_{-\tau_{\rm p}/2}^{\tau_{\rm p}/2} dt_2
  \int_{-c}^{c} ds_1 du_1 ds_2 du_2
  \frac{
  e^{i \Delta S/\hbar - (\tau_{s} + \tau_{u})/\tau_{\rm D}}}{
  (2 \pi \hbar)^2 t_{\rm enc,1}t_{\rm enc,2}}.
  \nonumber \\
  \label{eq:B}
\end{eqnarray}
Here $t_2$ parameterizes
the point at which the Poincar\'e surface of section for the reflection
trajectory of contact $2$ is taken, measured as the time needed to
travel between the Poincar\'e surfaces of sections for the
trajectories $1$ and $2$, see Fig.\ \ref{fig:6}. 
The time $t_{{\rm enc},i}$ is the
time during which trajectory $i$ remains within a phase space distance
$c$ from the periodic trajectory, $i=1,2$. 
The division by $t_{{\rm enc},i}$ cancels a
spurious contribution arising from the freedom to choose the
Poincar\'e surface of section along the trajectory. The times
$\tau_{s}$ and $\tau_{u}$ indicate the length of time over which the
trajectories $1$ and $2$ are correlated before and after getting
within a phase space distance $c$ of the closed loop, respectively,
see Fig.\ \ref{fig:6}.
The times $\tau_{j}$, $j=1,2,3,4$, indicate the duration of the four 
segments of uncorrelated propagation of the two pathes when they are
outside the encounter region. The classical action difference $\Delta S$ is
given in Eq.\ (\ref{eq:dS}) above.

We define the new integration variable
\begin{eqnarray}
  t_2' = t_2 - \lambda^{-1} \ln(c/|s_1|) + \lambda^{-1} \ln(c/|s_2|),
\end{eqnarray}
which is the time between the points where the 
trajectories $1$ and $2$
first come within a phase space distance $c$ from the periodic
trajectory.
Without loss of generality, we may assume that the phase space
coordinate $s_1$ is positive. However, we must keep the sign of the
phase space coordinate $s_2$ because it enters into the action
difference $\Delta S$ and into the correlation time $\tau_{s}$.
Making a variable change similar to that of Eq.\ (\ref{eq:varchg1}), 
\begin{eqnarray}
  s_i = c/\sigma_i,\ \ u_i = c x_i \sigma_i,\ \ i = 1,2,
\end{eqnarray}
we replace the double integration $ds_i du_i$, $i=1,2$, by a single
integration $dx_i$, $i=1,2$. This variable change cancels the
denominator $t_{{\rm enc},1} t_{{\rm enc},2}$ and adds a Jacobian
$(\lambda c^2)^2$, where $\lambda$ is the Lyapunov exponent, see Eq.\
(\ref{eq:F1calc}). 
We then find
\begin{eqnarray}
  B &=& \frac{N_1^2 N_2^2
  c^4 }{(2 \pi \hbar)^2 N^4} 
  \int d\tau_{\rm p} e^{-\tau_{\rm p}/\tau_{\rm D}} I(\tau_{\rm p}),
  \label{eq:BI}
\end{eqnarray}
with
\begin{eqnarray}
  I(\tau_{\rm p}) &=& 2 \lambda^2
  \int_{-\tau_{\rm p}/2}^{\tau_{\rm p}/2} dt_2'
  \int_{-1}^{1} dx_1 dx_2 \sum_{\pm}
  e^{-(\tau_{s}+\tau_{u})/\tau_{\rm D}}
  \nonumber \\ && \mbox{} \times
  \cos\left[r (x_1 \mp x_2) \right],
  \label{eq:I0}
\end{eqnarray}
where the sign $\pm$ is the sign of $s_2$ and
\begin{equation}
  r = \frac{c^2(1 - e^{-\lambda \tau_{\rm p}})}{\hbar}.
\end{equation}
The prefactor $2$ in Eq.\ (\ref{eq:I0}) arises
from fixing $s_1$ to be positive. Finally,
we change variables $u_1 = x_1$, $u_2 = \pm x_2$,
and $w = \pm e^{\lambda t_2'}$. The integration over $t_2'$ 
and summation over the sign $\pm$
of $s_2$ is then represented as the integral $\int
dw/\lambda |w|$, with $e^{-\lambda \tau_{\rm p}/2} < |w| < e^{\lambda \tau_{\rm p}/2}$,
\begin{eqnarray}
  I(\tau_{\rm p}) &=& 2 \lambda \int \frac{dw}{|w|}
  \int_{-1}^{1} du_1 du_2 
  \nonumber \\ && \mbox{} \times
  \cos[ r (u_1 - u_2)]
  e^{-(\tau_{s}+\tau_{u})/\tau_{\rm D}}.
  \label{eq:I0b}
\end{eqnarray}
The classical limit taken here corresponds to sending $r \to
\infty$, $\tau_{\rm D} \to \infty$, while keeping $\tau_{\rm
  E}/\tau_{\rm D} = (1/\lambda \tau_{\rm D}) \ln r$ fixed. Since
$I$ gets multiplied by $r^2 \tau_{\rm D}$, see Eq.\ (\ref{eq:BI})
above, we look for a leading contribution to $I$ of order $1/r^2
\tau_{\rm D}$. 

It is instructive to first consider the integral $I(\tau_{\rm p})$ without the
factor $\exp(-\tau_u/\tau_{\rm D})$. Since $\tau_{s}$ does not depend
on the integration variables $u_1$ and $u_2$, the integrals over
$u_1$ and $u_2$ are straightforward, and one finds
\begin{equation}
  I(\tau_{\rm p}) = \frac{8 \lambda}{r^2} \sin^2 r
  \int \frac{dw}{|w|} e^{-\tau_{s}/\tau_{\rm D}}.
\end{equation}
This result, however, is deceptive. The fact that $\sin^2 r \to 1/2$
for large $r$ is an artifact of our choice of the same
phase-space cut-off $c$ for the trajectories $1$ and $2$. The same
artifact appears when calculating $F_1^2$ for the contribution of type
a: when calculating $F_1$ we discarded a rapidly oscillating 
function $\propto \sin r$. However, the square of this rapidly
oscillating function has a nonzero average which should not be
retained in the final expression. We can avoid this problem by
taking slightly different phase-space cut-offs for the
trajectories $1$ and $2$, which amounts to
replacing $\cos(r u_1-r u_2))$ by $\cos(r_1 u_1 - r_2 u_2)$. We then
find
\begin{equation}
  I = 8 \lambda \int \frac{dw}{|w|} \frac{\sin r_1 \sin r_2}{r_1 r_2}
  e^{-\tau_{s}/\tau_{\rm D}}.
\end{equation}
In the classical
limit $r_1$, $r_2 \to \infty$ this is a fast oscillating
function and can be discarded. 
By symmetry, replacing the factor $\exp(-\tau_{s}/\tau_{\rm D})$ by
unity also gives a fast oscillating function which will be discarded 
from the final answer. Because of this, we may replace each factor
$\exp(-\tau_{s,u}/\tau_{\rm D})$ by $\exp(-\tau_{s,u}/\tau_{\rm
  D}) - 1$ in the integrand of Eq.\ (\ref{eq:I0b}). After this
replacement, we can restore $r_1 = r_2 = r$.

The fact that $I(\tau_{\rm p}) = 0$ without the factors
$\exp(-\tau_{s}/\tau_{\rm D})$ or $\exp(-\tau_{u}/\tau_{\rm D})$
signals the statistical independence of the conductance and the
density of states for a chaotic quantum dot with $\tau_{\rm E} \ll
\tau_{\rm D}$.\cite{kn:brouwer1997a} It is also crucial to ensure
unitarity. Without the
factors $\exp(-\tau_{s}/\tau_{\rm D})$ or $\exp(-\tau_{u}/\tau_{\rm D})$
the expression for $I(\tau_{\rm p})$ contains no reference to the phase space
distance between the trajectories $1$ and $2$ at the point that they
approach or leave the periodic trajectory, respectively. 
Without such reference, unitarity cannot be preserved, because there 
will be no difference between the case that the trajectories $1$ and $2$
originate/terminate at the same contact or not. With the factors
$\exp(-\tau_{s}/\tau_{\rm D})$ and $\exp(-\tau_{u}/\tau_{\rm D})$
unitarity is preserved, see the discussion at the end of this section.

At this point we should specify the times $\tau_{s}$ and
$\tau_{u}$. The escape of trajectories $1$ and $2$ is correlated
before arriving at the closed loop only if the stable phase space
coordinates $s_1$ and $s_2$ have the same sign, {\em i.e.}, if $w >
0$. In that case, the phase space distance between the two 
trajectories at the point of entrance of trajectory $1$ is 
$d = c |1 - \exp(-\lambda t_2')| = c|1 - 1/w|$ 
and the phase space distance between the
two trajectories at the point of entrance of trajectory $2$ is 
$d = c |\exp(\lambda t_2')-1| = c|w-1|$. Note that for $w$ close to $1$
(which is when escape correlations are relevant) the two phase space
distances are equal. For definiteness, and in order to have an
assignment that is symmetric under the exchange $1 \leftrightarrow 2$,
we use the largest of these two phase space distances in the following
considerations. Hence, if $w > 1$, we have $d = c(w-1)$, whereas $d =
c(1/w-1)$ if $0 < w < 1$. The correlation time (if positive), then, 
follows from setting $d e^{\lambda \tau_{s}} = c (b-1)$, where 
$e^{\lambda \tau_{\rm p}/2} > b> 1$ is a suitably chosen number of order 
unity,
\begin{equation}
  \tau_{s} = \left\{ \begin{array}{ll}
  \lambda^{-1} \ln[(b-1)/(w-1)] & \mbox{if $1 < w < b$}, \\
  \lambda^{-1} \ln[(b-1)/(1/w-1)] & \mbox{if $1/b < w < 1$}, \\
  0 & \mbox{otherwise}. \end{array} \right.
\end{equation}
The final results will be independent of the choice of the cut-off
$b$.

The definition of the correlation time $\tau_{u}$ is
similar. Proceeding as before, one finds that $\tau_{u}$ depends on
the product $z = w u_1/u_2$: this encodes both the time difference between
the exit points of trajectories $1$ and $2$ and the sign of the
unstable phase space coordinates at those points. There is one
subtlety when regarding escape for the exiting trajectories: if $|z| <
e^{-\lambda \tau_{\rm p}/2}$ or $|z| > e^{\lambda \tau_{\rm p}/2}$, $|z|$ has to be
brought back to the range $e^{-\lambda \tau_{\rm p}/2} < |z| < e^{\lambda
\tau_{\rm p}/2}$ by multiplication with the appropriate number of factors of
$e^{\pm \lambda \tau_{\rm p}}$. This procedure takes into account that the
trajectories $1$ and $2$ can make different numbers of revolutions
around the closed loop before exiting. (There was no such complication
when considering $\tau_{s}$, because the integration domain for $w$ is
$e^{-\lambda \tau_{\rm p}/2} < |w| < e^{\lambda \tau_{\rm p}/2}$.) 

Since $\tau_{s}$ and
$\tau_{u}$ appear in the combination $\exp(-\tau_{s,u}/\tau_{\rm D})$
only, we write $\exp(-\tau_{s}/\tau_{\rm D}) = f(w)$,
$\exp(-\tau_{u}/\tau_{\rm D}) = f(z)$. We then have $f(x) = 1$ if $x <
0$,
\begin{equation}
  f(x) = \left( \frac{x-1}{b-1} \right)^{1/\lambda \tau_{\rm D}}\ \
  \mbox{if $1 < x < b$},
\end{equation}
and $f(x) = 1$ if $b < x < e^{\lambda \tau_{\rm p}/2}$. 
Further, 
\begin{equation}
  f(x) = f(1/x)\ \ \mbox{and}\ \ f(x) = f(x e^{\lambda \tau_{\rm p}}).
\end{equation}
Note that the function $f$ is continuous
and piecewise differentiable. Also notice that $f = 1$ except in
the case of strong classical correlations between the trajectories $1$
and $2$.

With this choice of the function $f(z)$, we find that the integral
$I(\tau_{\rm p})$ reads
\begin{widetext}
\begin{eqnarray}
  I &=&
  \frac{4 \lambda}{r} \int \frac{dw}{w} [f(w) -1]
  \int_{0}^{1} du
  \left\{ 
  \int_{-|w|}^{|w|} dz [f(z) -1]
  \frac{\partial}{\partial z}
  \sin[u r(z/w - 1)]
  +
  \int_{-1/|w|}^{1/|w|} dz [f(z) -1]
  \frac{\partial}{\partial z}
  \sin[u r(z w - 1)]
  \right\}.
  \nonumber \\
  \label{eq:I1b}
\end{eqnarray}
Since $f(w) = 1$ if $w < 1/b$ or $w > b$, we can restrict the integration
range for $w$ to $1/b < w < b$. Performing a partial integration to
$z$, we have
\begin{eqnarray}
  I &=&
  -
  \frac{4 \lambda}{r} \int_{1/b}^{b} \frac{dw}{w} [f(w) - 1] 
  \int_0^1 du
  \left\{
  \int_{-w}^{w} dz \frac{\partial
  f(z)}{\partial z}
  \sin[u r(z/w - 1)]
  + \int_{-1/w}^{1/w} dz \frac{\partial
  f(z)}{\partial z}
  \sin[u r(z w - 1)] \right\}.
\end{eqnarray}
For the $z$ integration in the second line, there will be 
contributions from the regions $1/b < z < w$ and $e^{-n\lambda \tau_{\rm p}}/b < z <
e^{-n\lambda \tau_{\rm p}} b$, $n=1,2,\ldots$.
For the $z$ integration in the third line, there will be 
contributions from the regions $1/b < z < 1/w$, $e^{-n\lambda \tau_{\rm p}}/b < z <
e^{-n\lambda \tau_{\rm p}} b$, $n=1,2,\ldots$. 
In the region $1/b < z < b$ one has
\begin{eqnarray}
  \frac{\partial f(z)}{\partial z} &=&
  \frac{1}{(z-1)\lambda \tau_{\rm D}} f(z) \ \ \mbox{if $z > 1$},
  \nonumber \\
  \frac{\partial f(z)}{\partial z} &=&
  \frac{1}{z(z-1) \lambda \tau_{\rm D}} f(z) \ \ \mbox{if $z < 1$}.
\end{eqnarray}
Focusing on the integration range $1/b < z < b$, we then find
\begin{eqnarray}
  I &=& 
  \frac{8}{r \tau_{\rm D}}
  \int_{1}^{b} \frac{dw}{w} \int_{1}^{w} \frac{dz}{z-1}
  [f(w)-1] f(z)
  \int_0^1 du \left\{
  \sin[r u(1/zw-1)] - \sin[r u(z/w-1)] \right\}
  \nonumber \\ && \mbox{}
  + \frac{8}{r \tau_{\rm D}}
  \int_{1}^{b} \frac{dw}{w-1} \int_{1}^{w} \frac{dz}{z}
  [f(z)-1] f(w) 
  \int_0^1 du \left\{ \sin[r u(1/z w - 1)] +
  \sin[r u(z/w - 1)] \right\}.
\end{eqnarray}
Although these integrals can be evaluated 
for general $b$, the evaluation is simplest if $b-1 \ll 1$ (but still
$b-1$ of order unity). Writing $z = 1 + (b-1)\zeta/u$ and $w = 1 +
(b-1)\xi/u$ and expanding in $b-1$, one finds
\begin{eqnarray}
  I &=&
  - \frac{16(b-1)}{r \tau_{\rm D}}
  \int_0^1 d\xi \int_0^{1} \frac{d\zeta}{\zeta}
  \int_{\max(\xi,\zeta)}^{1} \frac{du}{u}
  [(\xi/u)^{1/\lambda \tau_{\rm D}} - 1]
  (\zeta/u)^{1/\lambda \tau_{\rm D}}
  \cos[r(b-1) \xi] \sin[r(b-1) \zeta].
  \label{eq:I3}
\end{eqnarray}
\end{widetext}
Performing the integral over $u$ followed by a partial integration
over $\xi$, the limit $r \to \infty$ at fixed ratio $\tau_{\rm
  E}/\tau_{\rm D} = (1/\lambda \tau_{\rm D}) \ln r$ can be taken. One
then finds
\begin{equation}
  I = 2\pi^2 (1 - r^{-2/\lambda \tau_{\rm D}})/r^2 \tau_{\rm D}.
\end{equation}

There is an alternative (and more intuitive) derivation of Eq.\ 
(\ref{eq:I3}) if we restrict our attention to 
interfering trajectories that arrive at and depart from the periodic 
trajectory at (classically) close phase space points from the very 
start of the calculation. This is the
situation drawn in the top panel of Fig.\ \ref{fig:6}. Instead of
choosing the Poincar\'e surfaces of section at an arbitrary point
during the encounter with the periodic trajectory, we may choose them
at the beginning and end of the encounter with the periodic trajectory
(indicated by the thin solid lines in the top panel of Fig.\
\ref{fig:6}). 
At the first (entrance) Poincar\'e surface of section, we use the
unstable phase space coordinate $u$ measured with respect to the
periodic trajectory, and the difference $s'$ of the stable phase space
coordinates of the trajectories $\alpha_1$ and $\beta_2$. 
(Since these trajectories depart from the periodic trajectory together,
their unstable phase space coordinate can be considered equal at this
Poincar\'e surface of section.)
Similarly, at the second (exit) Poincar\'e surface of section,
we use the common stable phase space coordinate $s$ of both trajectories,
measured with respect to the periodic trajectory, together with 
the difference $u'$ of the unstable phase space coordinates. The total time
the trajectory $\alpha_1$ spends near the periodic trajectory is
$\lambda^{-1} \ln(c/|u|) = \lambda^{-1} \ln(c/|s|)$, hence $|s| =
|u|$. In terms of these coordinates, $\exp(-\tau_{s}/\tau_{\rm D}) =
|s'/c(b-1)|^{1/\lambda \tau_{\rm D}}$ and $\exp(-\tau_{u}/\tau_{\rm D}) =
|u'/c(b-1)|^{1/\lambda \tau_{\rm D}}$.
Fixing the positions of the Poincar\'e surfaces of
section gives a Jacobian $\lambda c |u|$ and eliminates the factor
$t_{{\rm enc},1} t_{{\rm enc},2}$ from the denominator in Eq.\
(\ref{eq:B}).\cite{kn:mueller2005} 
Upon dividing all
coordinates by the phase space cut-off $c$, we then find
\begin{eqnarray}
  I &=&
  4 \lambda \int_{1-b}^{b-1} ds' du' \int_0^{1} du u
  (|s'/(b-1)|^{1/\lambda \tau_{\rm D}} - 1)
  \nonumber \\ && \mbox{} \times
  (|u'/(b-1)|^{1/\lambda \tau_{\rm D}} - 1)
  \cos[u r(s'-u')],
  \label{eq:I4}
\end{eqnarray}
where we added factors $2$ for the signs of $u$ and $s$. Rewriting the
integrals such that the integrations over $s'$ and $u'$ are all
between $0$ and $b-1$ and performing a partial integration to $s'$ we
find
\begin{eqnarray}
  I &=& -\frac{16}{\tau_{\rm D} r}
  \int_0^{b-1} \frac{ds'}{s'} du' 
  \int_0^1 du
  (|u'/(b-1)|^{1/\lambda \tau_{\rm D}} - 1)
  \nonumber \\ && \mbox{} \times
  |s'/(b-1)|^{1/\lambda \tau_{\rm D}}
  \sin(u r s') \cos(u r u').
\end{eqnarray}
Upon shifting variables $s' = \zeta (b-1)/u$ and $u' = \xi (b-1)/u$
one then arrives at Eq.\ (\ref{eq:I3}) above.

The $z$ integrations with $1/b < z e^{n \tau \lambda} < b$ with $n =
1$, $2$, \ldots\ do not give a contribution to $I(\tau_{\rm p})$ in the limit
$r \to \infty$. This can be seen by noting that all oscillating 
integrals all contain fast oscillating phases proportional to $r(1 - 
e^{-n \lambda \tau_{\rm p}})$.

Putting everything together, we find
\begin{equation}
  B = \frac{N_1^2 N_2^2}{2 (N_1+N_2)^4 \tau_{\rm D}} \int d\tau_{\rm p}
  \frac{e^{-\tau_{\rm p}/\tau_{\rm D}}}{(1 - e^{-\lambda \tau_{\rm p}})^2}
  (1 - e^{-2 \tau_{\rm E}/\tau_{\rm D}}).
\end{equation}
Setting a lower cut-off for the $\tau_{\rm p}$-integration at 
$\tau_{\rm p} \gtrsim
1/\lambda$ and taking the limit $\tau_{\rm D} \lambda \gg 1$, we
finally arrive at the simple result
\begin{equation}
  B = \frac{N_1^2 N_2^2}{2 (N_1+N_2)^4}(1 - e^{-2 \tau_{\rm E}/\tau_{\rm D}}).
\end{equation}
Substitution into Eq.\ (\ref{eq:varGF}) then gives
\begin{equation}
  \mbox{var}\, G = 2 \left( \frac{2 e^2}{h} \right)^2
  \frac{N_1^2 N_2^2}{(N_1+N_2)^4}.
  \label{eq:varG}
\end{equation}

Equation (\ref{eq:varG}) is the main result of this article. The
variance of the conductance is independent of the Ehrenfest time.
Equation (\ref{eq:varG}) was derived for a quantum dot without an
applied magnetic field. With a magnetic field strong enough to 
fully break time-reversal symmetry $\mbox{var}\, G$ is reduced by a
factor two.

The most remarkable feature of Eq.\ (\ref{eq:varG}) is that the
conductance fluctuations survive in the limit $\tau_{\rm E} \gg
\tau_{\rm D}$. The particular classical trajectories that give rise to
conductance fluctuations in this limit can be identified by inspection
of the calculation above. The constant term in Eq.\ (\ref{eq:varG})
can be traced back to the lower limit on the $u$ integration in Eq.\
(\ref{eq:I3}) which, in turn, results from trajectories that wind
many times around the periodic trajectory.
Although such
trajectories must spend a time $\gtrsim \tau_{\rm E}$ inside the
quantum dot in order to contribute to the conductance fluctuations, their
survival probability depends on the period $\tau_{\rm p}$ of the periodic
trajectory only, not on the actual time they spend inside
the quantum dot.

Instead of calculating the conductance variance through the covariance
of the reflection off the two contacts, one can also directly
calculate the variance of the conductance or the variance of the
reflection. As in the case of the calculation of the quantum
correction to the average conductance, there are two types of
contributions to the fluctuations: interfering trajectories for which
all encounters lie within the interior of the quantum dot and interfering
trajectories for which at least one encounter touches the lead
opening. The calculation of the first type, all encounters inside
the quantum dot, proceeds in essentially the same was as the calculation of 
the reflection covariance outlined above. Below, we discuss how the
above calculations should be modified to include the 
second type, for which one or more encounters 
touch the lead opening. We find that, once encounters that touch the
lead opening are taken into account, unitarity is obeyed for
contributions of type a and b separately.

For trajectories of type a, there are two encounters that each can be
close to a lead opening. For each encounter, however, the
configuration of trajectories is 
identical to that of coherent backscattering calculation. Repeating
the steps of the last part of the previous section, verification of
unitarity is immediate. 

\begin{figure}[t]
\epsfxsize=0.8\hsize
\epsffile{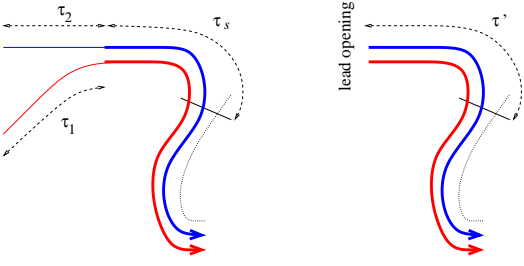}
\caption{
  \label{fig:7}
Comparison of a part of a type b encounter that fully resides inside
the quantum dot (left) and an encounter that touches to lead opening
(right). The relevant segments of the (short) trajectories $\alpha_1$
and $\beta_2$ are shown solid, the
periodic trajectory is shown dotted.}
\end{figure}

For trajectories of type b, a schematic drawing showing the difference
between an encounter that fully resides inside the quantum dot and an
encounter that touches the lead opening is shown in Fig.\
\ref{fig:7}. For an encounter that touches the lead opening, 
one replaces the phase space coordinate differences $s'$
or $u'$ in Eq.\ (\ref{eq:I4}) by $v$, where $v$ represents the
perpendicular component of the momentum in the lead opening. As
in Sec.\ \ref{sec:2}, 
we normalize $v$ such that the volume element
in phase space is $du dv$. With this choice of phase space
coordinates, the action difference for 
two pairs of trajectories involved in an
encounter that resides inside the quantum dot is the same as the action
difference for two pairs of trajectories for which one of the
encounter extends to the lead opening, up to the substitution $s'
\to v$ or $u' \to v$. 
One then obtains the
contribution of encounters that touch the lead opening by making the 
replacement
\begin{eqnarray}
  \lefteqn{\frac{N_i}{N_1+N_2}
  \left(
  \prod_{j=1}^{2} \int_0^{\infty}
  \frac{d\tau_{j}}{\tau_{\rm D}}
  e^{-\tau_{j}/\tau_{\rm D}}\right)
  e^{-\tau_{s,u}/\tau_{\rm D}}} \nonumber \\
  &\to&
  \int_0^{\tau_{v}} \frac{d\tau'}{\tau_{\rm D}} e^{-\tau'/\tau_{\rm
  D}}
  ~~~~~~~~~~~~~~~~~~~~~~~
\end{eqnarray}
in the expression for the correlator with encounters that reside
inside the quantum dot only. Here $\tau'$ is the travel time between the
lead opening and the point where the two trajectories get close to the
periodic trajectory and $\exp(-\tau_v/\tau_{\rm D}) =
|v/c(b-1)|^{1/\lambda \tau_{\rm D}}$.
The $\tau'$ integration gives a factor
$1 - e^{-\tau_{v}/\tau_{\rm D}}$, so that the remaining integrals
are the same as for the case that all encounters are inside the quantum dot
and one verifies that unitarity is obeyed for type b encounters as well.

\section{Time dependence}
\label{sec:4}

Although the calculation of the previous section answers the question
which trajectories are responsible for the conductance fluctuations in
the limit of large Ehrenfest times --- trajectories that wind many
times around a certain periodic trajectory ---, it does not tell us
how long these trajectories spend inside the quantum dot. That question can
be answered by adding an imaginary term to the energy. Such an
imaginary term gives rise to an additional exponential decay
$\exp(-t_{\alpha}/2 \tau_{\rm abs})$ for each trajectory $\alpha$,
where $t_{\alpha}$ is the total duration of trajectory $\alpha$ and
$\tau_{\rm abs}$ is the corresponding absorption time. A minimal time
needed for quantum interference shows up through an exponential dependence
on $1/\tau_{\rm abs}$.
In the calculations below, we keep the ratios $\tau_{\rm E}/\tau_{\rm
  D}$ and $\tau_{\rm E}/\tau_{\rm abs}$
fixed while taking the classical limit $\hbar \to 0$.

The addition of absorption has been used in the numerical simulations
of Refs.\ \onlinecite{kn:tworzydlo2004c,kn:rahav2005,kn:rahav2006} as
a diagnostic tool to investigate the microscopic mechanism of quantum
interference corrections. Reference \onlinecite{kn:rahav2006} found
that the difference between weak localization and conductance
fluctuations not only concerned the dependence of their magnitude on
the Ehrenfest time --- exponential decay versus independence of the
Ehrenfest time ---, but also the minimal dwell time of trajectories
that contribute to quantum interference. According to the numerical
simulations, the minimal dwell time for weak localization is $2
\tau_{\rm E}$, whereas the minimal dwell time of trajectories
contributing to conductance fluctuations was found to be $\tau_{\rm
E}$. Below we show that this observation is fully consistent with the
semiclassical theory.

We first consider weak localization. With the imaginary
term added to the energy, one finds that the quantum correction to the
transmission is given by
\begin{eqnarray}
  \delta T &=& \frac{N_1 N_2}{(N_1+N_2)^2}
  \left( \prod_{j=1}^{2} \int_0^{\infty} \frac{d\tau_{j}}{\tau_{\rm
  D}}
  e^{-\tau_{j}/\tau_{\rm D} - \tau_{j}/\tau_{\rm abs}} \right)
  \nonumber \\ && \mbox{} \times
  \int d\tau_3 e^{-\tau_3/\tau_{\rm D} - \tau_3/\tau_{\rm abs}}
  \nonumber \\ && \mbox{} \times
  \int_{-c}^{c} ds du \frac{e^{i s u/\hbar - t_{\rm enc}/\tau_{\rm D}
  - 2 t_{\rm enc}/\tau_{\rm abs}}}{2 \pi \hbar t_{\rm enc}}.
\end{eqnarray}
Note the factor two in front of $t_{\rm enc}/\tau_{\rm abs}$ in
the survival probability during the encounter. This factor two arises
because
trajectories that contribute to weak localization travel the encounter
region twice.
Performing the integrations as in Sec.\ \ref{sec:2}, one finds
\begin{equation}
  \delta T = -\frac{N_1 N_2}{(N_1+N_2)^2}
  \frac{\left(1 + 2 \tau_{\rm D}/\tau_{\rm abs} \right)}{
  \left(1 + \tau_{\rm D}/\tau_{\rm abs} \right)^{3}
  }e^{-\tau_{\rm E}/\tau_{\rm D} - 2 \tau_{\rm E}/\tau_{\rm
  abs}}.  
\end{equation}
Similarly, for reflection one finds
\begin{eqnarray}
  \delta R_i &=& \frac{N_1 N_2}{(N_1+N_2)^2}
  \frac{\left(1 + 2 \tau_{\rm D}/\tau_{\rm abs} \right)}{
  \left(1 + \tau_{\rm D}/\tau_{\rm abs} \right)^{3}
  }e^{-\tau_{\rm E}/\tau_{\rm D} - 2 \tau_{\rm E}/\tau_{\rm
  abs}}
  \nonumber \\ && \mbox{} +
  \frac{N_i}{N_1 + N_2} \frac{(\tau_{\rm D}/\tau_{\rm abs})^2}{\left(1
  + \tau_{\rm D}/\tau_{\rm abs} \right)^{3}}
  e^{-\tau_{\rm E}/\tau_{\rm D} - 2 \tau_{\rm E}/\tau_{\rm
  abs}}.\nonumber \\
\end{eqnarray}
In the limit $\tau_{\rm E} \ll \tau_{\rm D}$, $\tau_{\rm abs}$ these
results agree with those obtained from random matrix
theory.\cite{kn:brouwer1997c} Note that $\delta R_i$ no longer equals 
$-\delta T$, $i=1,2$, in the presence of absorption.

The exponential dependence $\propto \exp(-2 \tau_{\rm E}/\tau_{\rm
abs})$ indicates that the minimal duration of a trajectory that
contributes to weak localization is $2 \tau_{\rm E}$: a self-encounter
lasts one Ehrenfest time and each trajectory to weak localization passes
through the encounter region twice. The fact that
the dwell time dependence $\propto \exp(-\tau_{\rm E}/\tau_{\rm
D})$ does not have this factor two is a result of classical 
correlations between the two segments of the trajectory that 
pass through the encounter region.\cite{kn:rahav2005,kn:heusler2006} 
Both effects --- the
minimal time $2 \tau_{\rm E}$ needed for weak localization and the
exponential suppression $\propto \exp(-\tau_{\rm E}/\tau_{\rm
D})$ are consistent with the numerical simulations of
Refs.\ \onlinecite{kn:rahav2005,kn:rahav2006}.

For the conductance fluctuations we treat the contributions of
trajectories of types a and b separately. We limit our discussion to
the covariance $\cov(R_1,R_2)$ of the reflections off the two point
contacts. Including absorption into the covariance contribution 
$A$ of trajectories of type a, one finds
\begin{equation}
  A = \frac{(N_1 N_2)^2 }{(N_1+N_2)^4}
  \frac{\left(1 + 2 \tau_{\rm D}/\tau_{\rm abs} \right)^2}{
  (1 + \tau_{\rm D}/\tau_{\rm abs} )^{6}
  }e^{-2 \tau_{\rm E}/\tau_{\rm D} - 4 \tau_{\rm E}/\tau_{\rm
  abs}}.  
\end{equation}

In order to calculate the contribution of trajectories of type b, we
need to calculate the integral $I(\tau_{\rm p})$ in the presence of
absorption,
\begin{eqnarray}
  I(\tau_{\rm p}) &=& 2 \lambda e^{-\tau_{\rm p}/\tau_{\rm abs}}
  \int \frac{dw}{|w|} \int_{-1}^{1} du_1 du_2
  |u_1 u_2|^{1/\lambda \tau_{\rm abs}}
  \nonumber \\ && \mbox{} \times
  \cos(r_1 u_1 - r_2 u_2) e^{-(\tau_s + \tau_u)(1/\tau_{\rm D} +
  2/\tau_{\rm abs})}. \nonumber \\
\end{eqnarray}
Here we used that the time the short trajectories spend near the periodic
trajectory is $\lambda^{-1} \ln(1/|u_i|)$, $i=1,2$, and that
the time spent by the long trajectory is $\tau_{\rm p} + \lambda^{-1}
\ln(1/|u_i|)$. 
As before, 
we first calculate $I$ without the exponential factors
involving $\tau_s$ or $\tau_u$. We then find
\begin{equation}
  I(\tau_{\rm p}) =  
  \frac{4 \pi^2 \tau_{\rm p}}{r^2 \tau_{\rm abs}^2}
  e^{-\tau_{\rm p}/\tau_{\rm abs}-2 \tau_{\rm E}/\tau_{\rm abs}}. 
\end{equation}
Replacing one exponential factor $\exp[-\tau_{s,u}(1/\tau_{\rm D} +
2/\tau_{\rm abs})]$ by $\exp[-\tau_{s,u}(1/\tau_{\rm D} +
2/\tau_{\rm abs})]-1$ while still setting $\tau_{u,s} = 0$ in the
other exponential factor,
one finds no significant contribution to $I$. Hence, the remaining
contribution to $I$ can be calculated by replacing both factors
$\exp[-\tau_{s,u}(1/\tau_{\rm D} + 2/\tau_{\rm abs})]$ by 
$\exp[-\tau_{s,u}(1/\tau_{\rm D} + 2/\tau_{\rm abs})] - 1$.
The result then follows from Eq.\ (\ref{eq:I3}) after
addition of a factor $u^{2/\lambda \tau_{\rm abs}}$ and replacement of
$1/\tau_{\rm D}$ by $1/\tau_{\rm D} + 2/\tau_{\rm abs}$.
Performing the integrals as
described in Sec.\ \ref{sec:3}, we find
\begin{eqnarray}
  I(\tau_{\rm p}) &=& \frac{4 \pi^2 \tau_{\rm p}}{r^2 \tau_{\rm abs}^2}
  e^{-\tau_{\rm p}/\tau_{\rm abs}-2 \tau_{\rm E}/\tau_{\rm abs}}
  \nonumber \\ &&
  \mbox{} +
  \frac{  2 \pi^2
  (1/\tau_{\rm D} + 2/\tau_{\rm abs})^2}{r^2 (1/\tau_{\rm D} +
  1/\tau_{\rm abs})} e^{-\tau_{\rm p}/\tau_{\rm abs}}
  \nonumber \\ && \mbox{} \times
  \left[  e^{-2 \tau_{\rm E}/\tau_{\rm abs}}
  - e^{-2 \tau_{\rm E}/\tau_{\rm D} - 4 \tau_{\rm
  E}/\tau_{\rm abs}}\right].
\end{eqnarray}
Adding the contributions of trajectories of types a and b, the 
final result becomes
\begin{eqnarray}
  \cov(R_1,R_2) &=&
  \frac{(N_1 N_2)^2}{(N_1+N_2)^4}
  e^{-2 \tau_{\rm E}/\tau_{\rm abs}}
  \\ && \mbox{} \times
  \frac{(1 + 2 \tau_{\rm D}/\tau_{\rm abs})^2
  + 2 (\tau_{\rm D}/\tau_{\rm abs})^2}{(1 + \tau_{\rm
  D}/\tau_{\rm abs})^6}. \nonumber
\end{eqnarray}
The limit $\tau_{\rm E}/\tau_{\rm D} \to 0$ agrees with the result
obtained from random matrix theory.\cite{kn:brouwer1997c} 
The exponential decay $\propto \exp(-2\tau_{\rm E}/\tau_{\rm D})$
corresponds to a minimal dwell time $\tau_{\rm E}$ for trajectories
that contribute to conductance fluctuations. This is consistent with
the numerical simulations of Refs.\ \onlinecite{kn:tworzydlo2004c} and
\onlinecite{kn:rahav2006}. The same conclusion remains true for other
correlators of transmissions and reflections (which need not be equal
to $\mbox{cov}\,(R_1,R_2)$ in the presence of absorption).

\section{Conclusion}

In the previous sections, we have calculated the Ehrenfest-time 
dependence of 
the weak localization correction and the conductance fluctuations of
a ballistic quantum dot with ideal point contacts, using a
trajectory-based semiclassical theory. We find that the Ehrenfest-time
dependences of weak localization and conductance fluctuations are
remarkably different: whereas weak localization is suppressed
exponentially if the Ehrenfest time $\tau_{\rm E}$ is much larger than
the mean dwell time $\tau_{\rm D}$, the conductance fluctuations are
independent of the ratio $\tau_{\rm E}/\tau_{\rm D}$. Our calculation
explains the numerical simulations of the conductance fluctuations in
Refs.\ \onlinecite{kn:tworzydlo2004,kn:jacquod2004}.

The numerical observation of Ehrenfest-time independent conductance
fluctuations was remarkable, because trajectories need to remain
inside the quantum dot during at least a time $\tau_{\rm E}$ if they are
to contribute to quantum interference corrections, so that one expects
that interference corrections to the conductance should disappear if
$\tau_{\rm E} \gg \tau_{\rm D}$. Our calculation shows that the latter
expectation is not always born out. The Ehrenfest-time independent 
contribution to the conductance fluctuations we find here arises from 
pairs of trajectories that wind many times around a periodic 
trajectory. On the one hand, 
such configurations of interfering trajectories have 
sufficient duration to allow for small action differences and thus
provide a significant quantum interference correction. On the other
hand, they are not 
as much affected by classical escape into the leads: The survival
probability depends on the period of the periodic trajectory involved,
not on the Ehrenfest time.

In Refs.\ \onlinecite{kn:tworzydlo2004} and
\onlinecite{kn:jacquod2004} the
`effective random matrix theory' of Silvestrov {\em et
al.}\cite{kn:silvestrov2003} was used to
explain the numerical observation of Ehrenfest-time independent
conductance fluctuations. According to the effective random matrix
theory, phase space is separated into a `classical part',
corresponding to all trajectories with dwell time less than the
Ehrenfest time, and a `quantum part', which has all trajectories with
dwell time larger than $\tau_{\rm E}$. The wave nature of the
electrons plays no role in the classical part of phase space, whereas
the quantum part of phase space is described using random matrix
theory. The effective random matrix theory was able to correctly
describe the Ehrenfest-time dependence of shot
noise\cite{kn:tworzydlo2003} and the density
of states of an Andreev quantum
dot,\cite{kn:silvestrov2003,kn:beenakker2004,kn:goorden2005} but it
missed the
Ehrenfest-time dependence of weak localization.\cite{kn:rahav2005} 
According to our semiclassical theory, the effective random matrix 
theory not only gave the
correct magnitude of the universal conductance fluctuations, but it
also gives the right dependence on an imaginary 
potential.\cite{kn:tworzydlo2004c} While it is understood 
that the
effective random matrix theory is not a comprehensive theory of the
Ehrenfest time dependence of quantum transport, it remains an interesting
question to classify which transport phenomena are described by this
phenomenological theory and which are not.


\acknowledgments

We especially thank Petr Braun, Fritz Haake, Stefan Heusler, and
Sebastian M\"uller for the collegial manner in which they alerted 
us to a mistake in an earlier version of this manuscript and for
correspondence. The calculation of Sec.\ \ref{sec:4} was motivated by
discussions with Alexander Altland. We thank Igor Aleiner, Alexander
Altland, Carlo
Beenakker, and Philippe Jacquod for discussions and encouragement.
This work was supported by the NSF under grant no.\ DMR 0334499 and 
by the Packard Foundation.


\end{document}